\documentclass[aps,prc,showpacs,twocolumn,superscriptaddress]{revtex4-1}
\usepackage{amsmath,amsfonts,amssymb,amsthm,float,mathastext, array, booktabs, tabularx, mathtools}
\usepackage{booktabs}
\usepackage{graphicx}
\graphicspath{{fig/}}
\usepackage{siunitx}
\usepackage{hyperref}
\hypersetup{hidelinks,colorlinks=false,breaklinks=true,urlcolor=ocre}
\usepackage{dcolumn}% Align table columns on decimal point
\usepackage{bm}% bold math
\usepackage[english]{babel}
\usepackage[autostyle]{csquotes}
\usepackage{color}
\begin{document}
\preprint{APS/123-QED}

\title{Evidence of cluster dipole states in germanium detectors operating at temperatures below 10 K}% Force line breaks with \\

\author{D.-M. Mei} 
 \email{Corresponding author.\\Email: Dongming.Mei@usd.edu}
\affiliation{Department of Physics, The University of South Dakota, Vermillion, SD 57069, USA}
\author{R. Panth}
\affiliation{Department of Physics, The University of South Dakota, Vermillion, SD 57069, USA}
\author{K. Kooi}
\affiliation{Department of Physics, The University of South Dakota, Vermillion, SD 57069, USA}
\author{H. Mei}
\affiliation{Department of Physics, The University of South Dakota, Vermillion, SD 57069, USA}
\author{S. Bhattarai}
\affiliation{Department of Physics, The University of South Dakota, Vermillion, SD 57069, USA}
\author{M. Raut}
\affiliation{Department of Physics, The University of South Dakota, Vermillion, SD 57069, USA}
\author{P. Acharya}
\affiliation{Department of Physics, The University of South Dakota, Vermillion, SD 57069, USA}
\author{G.-J. Wang}
\affiliation{Department of Physics, The University of South Dakota, Vermillion, SD 57069, USA}

\date{\today}% It is always \today, today,%  but any date may be explicitly specified

\begin{abstract}
By studying charge trapping in germanium (Ge) detectors operating at temperatures below 10 K, we demonstrate for the first time that the formation of cluster dipole states from residual impurities is responsible for charge trapping. Two planar detectors with different impurity levels and types are used in this study. When drifting the localized charge carriers created by $\alpha$ particles from the top surface across a detector under a lower bias voltage, significant charge trapping is observed when compared to operating at a higher bias voltage. The amount of charge trapping shows a strong dependence on the type of charge carriers. Electrons are trapped more than holes in a p-type detector while holes are trapped more than electrons in a n-type detector. When both electrons and holes are drifted simultaneously using the widespread charge carriers created by $\gamma$ rays inside the detector, the amount of charge trapping shows no dependence on the polarity of bias voltage. 
 
\end{abstract}
%\pacs{29.40.Mc, 24.10.-i, 29.85.Fj, 13.75.-n}% PACS, the Physics and Astronomy
\keywords{Suggested keywords}%Use showkeys class option if keyword
\maketitle
%\tableofcontent

  Low-background germanium (Ge) crystal detectors are a well-accepted technology in the search for dark matter (DM)~\cite{cogent, cdms, supercdms, edelweiss, cdex} and neutrinoless double-beta (0$\nu\beta\beta$) decay~\cite{hvkk, igex, gerda, majorana}. The next generation of ton-scale experiments aims to achieve: (a) an extremely low-energy threshold, below 100 eV of nuclear recoils, to detect low-mass DM particles~\cite{essig, ahmed, mei} and (b) an extremely low-level background rate, 0.01 events/ton/year in the region of interest (ROI)~\cite{legend}, for 0$\nu\beta\beta$ decay. This requires large-size Ge crystals (2 to 4 kg) with sufficient purity for developing detectors with: (i) internal charge or phonon amplification for DM~\cite{mei, ahmed} and (ii) capable of differentiating a two-electron signature from 0$\nu\beta\beta$ decay and Compton scatters from background events~\cite{legend}. Both require the charge trapping in Ge detectors to be well-understood in order to achieve best energy resolution. While the charge trapping is relatively well-understood for the detectors operating at temperatures around~\cite{gerda, majorana} 77 K and milli-Kelvin~\cite{supercdms, edelweiss}, very little is understood for the charge trapping in the detectors operating around the liquid helium temperature.
  
  The residual impurities (donors or acceptors) in Ge freeze out from the conduction or valence band to localized states, forming electric dipoles or neutral states D$^{0}$ and A$^{0}$ around liquid helium temperature ($\sim$4 K)~\cite{abak1, abak2, gsad, dven}. As a result, the density of free charge carriers decreases exponentially as temperature decreases~\cite{nea, str}. When the temperature approaches below 10 K, most of the donor (or acceptor) atoms remain unionized. The fifth electrons (or the empty holes) are confined to the donor (or the acceptor) atoms. The range of the confinement is determined by the Onsager radius~\cite{sager, onsager, ual,mei1}, R=$\frac{1}{4\pi\epsilon\epsilon_{0}k_{B}T}$, where $\epsilon$ = 16.2 is the relative permittivity for Ge, $\epsilon_{0}$ is the permittivity of free space, $k_{B}$ is the Boltzmann constant, and $T$ is temperature. This Onsager radius can be much larger than the size of donor or acceptor atom at temperatures below 10 K. Consequently, the fifth electrons (or the empty holes) tied to the donor (or acceptor) atoms can be thermally separated from the cores of the atoms. This separation of positive and negative charges forms an electric dipole. Since the orbital position of bound electrons or holes can be thermally shifted away from their original orbits, the bound electrons or holes are in the excited states. Therefore, the formed dipole states are thermally excited, labelled as $D^{0^{*}}$ (or $A^{0^{*}}$), which is different from the ground state of $D^{0}$ (or $A^{0}$).  
 
\begin{figure} [htbp]
  \centering
  \includegraphics[clip,width=0.8\linewidth]{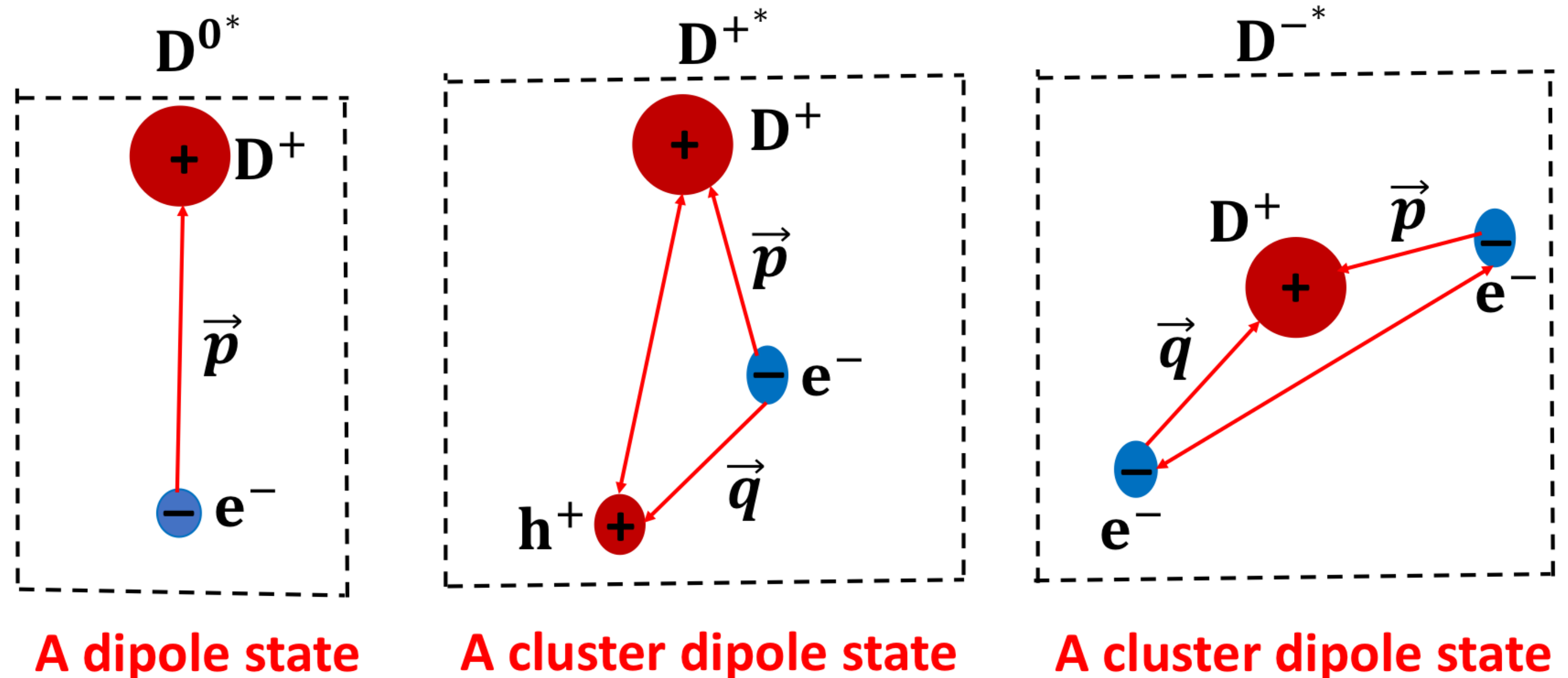}
   \includegraphics[clip,width=0.8\linewidth]{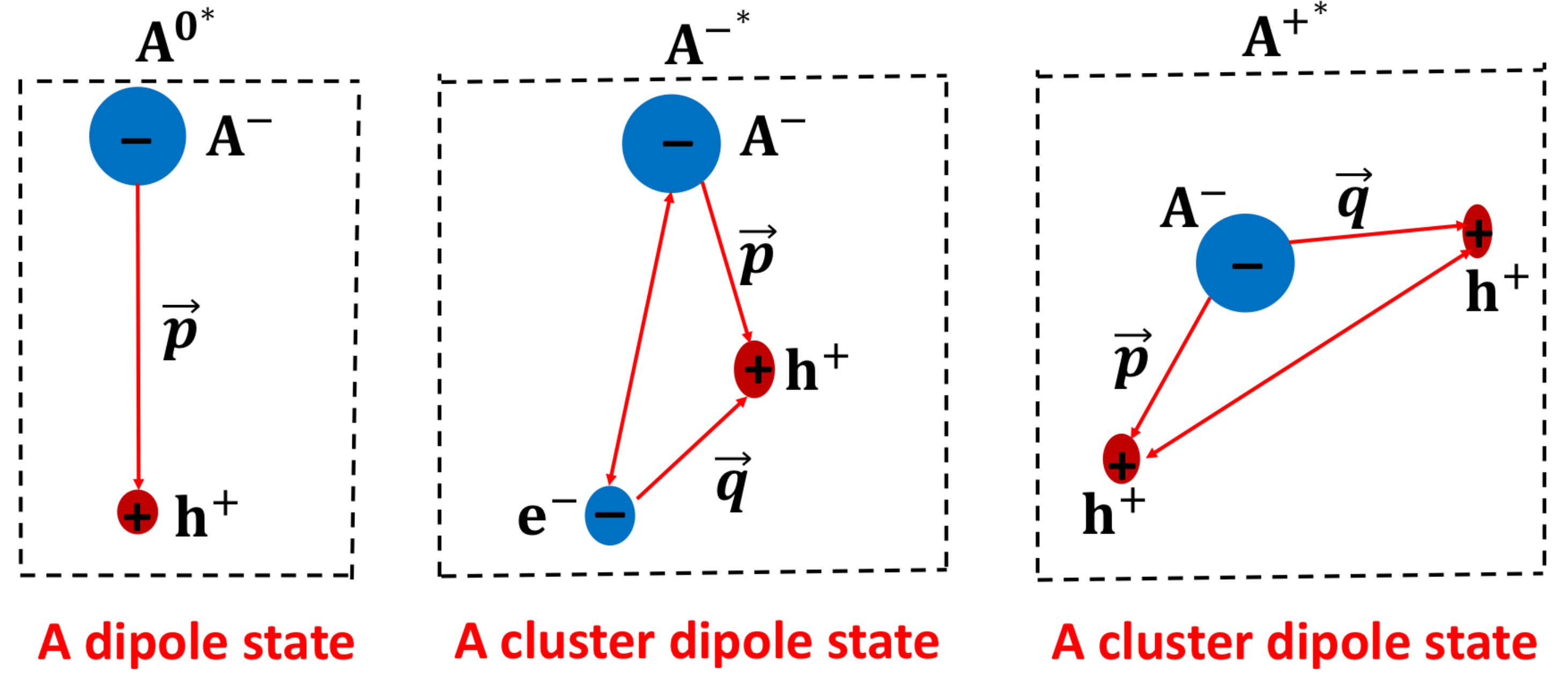}
  \caption{The sketch of the processes involved in the formation of the excited dipole states and the cluster dipole states in an n-type (upper) and a p-type (lower) Ge detector operated at temperatures below 10 K, where $\vec{p}$ and $\vec{q}$ are the corresponding dipole moments.}
  \label{fig:f1}
\end{figure}

 Subsequently, the formed dipole states can trap charge through Coulomb attraction to form cluster dipole states. 
 Figure~\ref{fig:f1} depicts the formation of excited dipole states and cluster dipole states in n-type Ge or p-type Ge, respectively. The excited dipole states can be shallow traps or deep traps~\cite{lax, mei1}. Shallow traps form cluster dipole states where the charge can be released through phonon excitation~\cite{lax, mei1}. While deep traps produce single space-charged states because positive and negative charges are recombined~\cite{lax, mei1} through capture. The probability of shallow traps is much larger than that of deep traps according to the Lax model~\cite{lax}. 

 Within an excited dipole state, the positively charged donor ion (or negatively charged acceptor ion) is deeply confined by the lattice deformation potential~\cite{jba}, the allowable phase space for trapping charge carriers is smaller than that of bound electrons (or holes), which can move within the range confined by the Onsager radius. Thus, it is expected that the probability of forming $D^{+^{*}}$ states (or $A^{-^{*}}$ states) is larger than that of $D^{-^{*}}$ states (or $A^{+^{*}}$ states) in an n-type detector (or a p-type detector). This is to say that holes are trapped more severely compared to electrons in an n-type detector and electrons are trapped more than holes in a p-type detector. Note that although the trapping of charge carriers by excited dipole states to form cluster dipole states has much in common with binding of the carriers into excitons, the formation processes are very different~\cite{exciton}.                  

 We report here the results of the spectral analysis of charge trapping of $D^{0^{*}}$ (or $A^{0^{*}}$) states in an n-type (or a p-type) Ge planar detector operated at cryogenic temperature of 5.2 K. The measured energy deposition of the 5.3 MeV $\alpha$ from an $^{241}$Am source is used as the detecting mechanism. The analysis is compared to the same detector operated at 77.8 K. The charge trapping mechanism is demonstrated using the localized charge carriers created by $\alpha$ particles from the top of the surface across the detector under a lower bias voltage. The dependence of charge trapping on the type of charge carriers (electrons vs holes) in a given detector is shown when the detector is operated with different polarities of bias voltage (positive vs negative). The widespread charge carriers created by 661.7 keV $\gamma$ rays from a $^{137}$Cs source are used to verify the trapping mechanism.   
 
\begin{figure}[htbp]
  \centering
  \includegraphics[clip,width=0.9\linewidth]{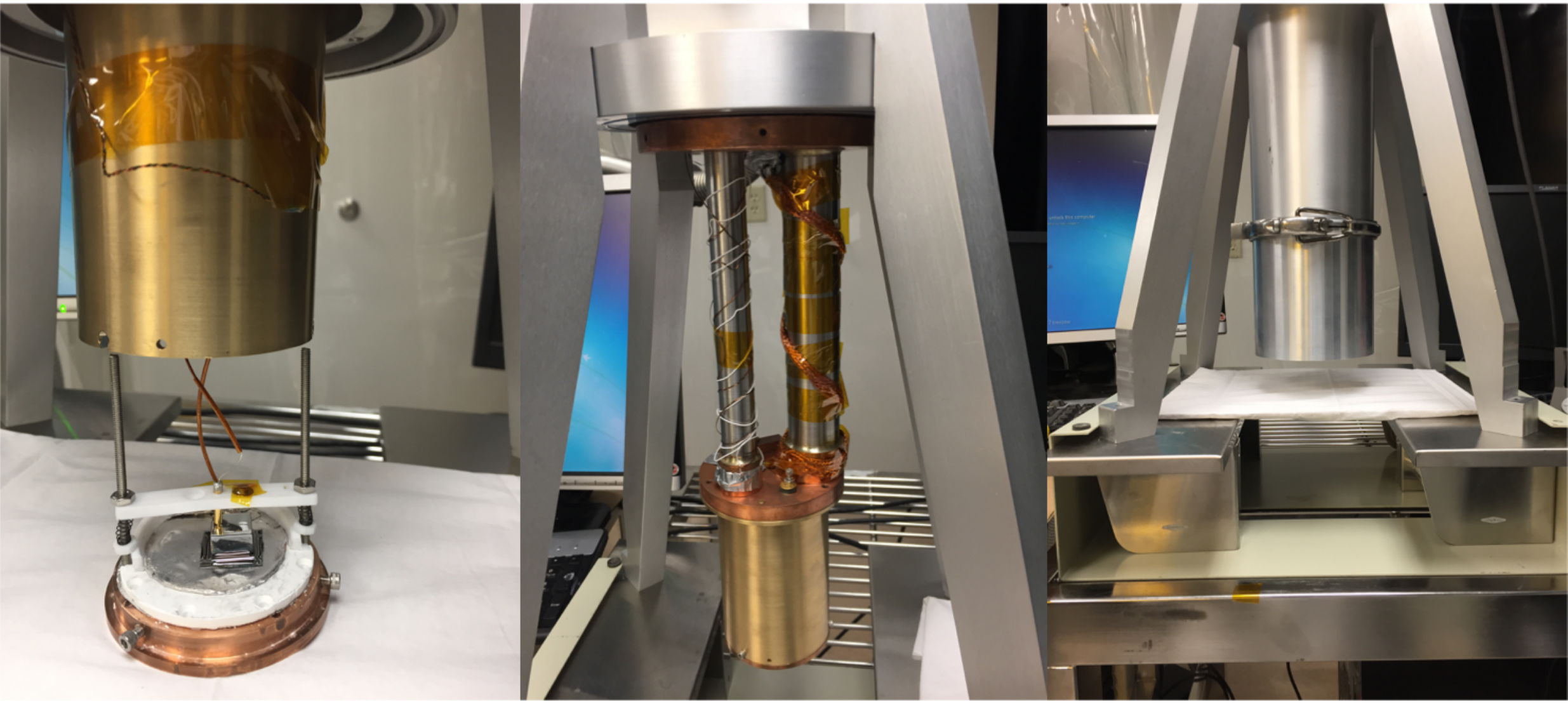}
  \caption{Shown is the experimental setup for measuring energy deposition of the 5.3 MeV $\alpha$ particles from a $^{241}Am$ source and the 661.7 keV $\gamma$ rays from a $^{137}$Cs source. Left: a planar detector is mounted in the cryostat; Middle: the detector is enclosed in the house with temperature sensors; Right: the enclosed cryostat. }
  \label{fig:f3}
\end{figure}

 Both n-type and p-type Ge crystals grown at the University of South Dakota~\cite{wang1, wang2, wang3} were fabricated into thin-contact planar detectors~\cite{mang, wei1, wei2}. A layer of amorphous Ge (a-Ge) with a thickness of $\sim$600 nm was coated on the Ge crystal surface as the electrical contact to block the surface charges~\cite{wei1, wei2}. On the surface of the a-Ge, a thin aluminum (Al) layer ($\sim$100 nm) was deposited by using a sputtering machine to increase conductivity. The detector is mounted into a cryostat  and cooled down to about 5.2 K using a pulse tube refrigerator. Figure~\ref{fig:f3} shows the experimental setup. There are two temperature sensors inside the detector house. One is mounted at the bottom of a copper plate on which the detector is set on a thin indium foil. The other one is set on the top of another copper plate, which is close to the top surface of the detector. Thus, the detector is set between two temperature sensors. Since the measured temperatures from the two sensors are always different from each other by less than 0.5 K, this configuration ensures the detector temperature is measured with an accuracy of 0.5 K.  
 
\begin{figure}[htbp]
  \centering
  \includegraphics[clip,width=0.8\linewidth]{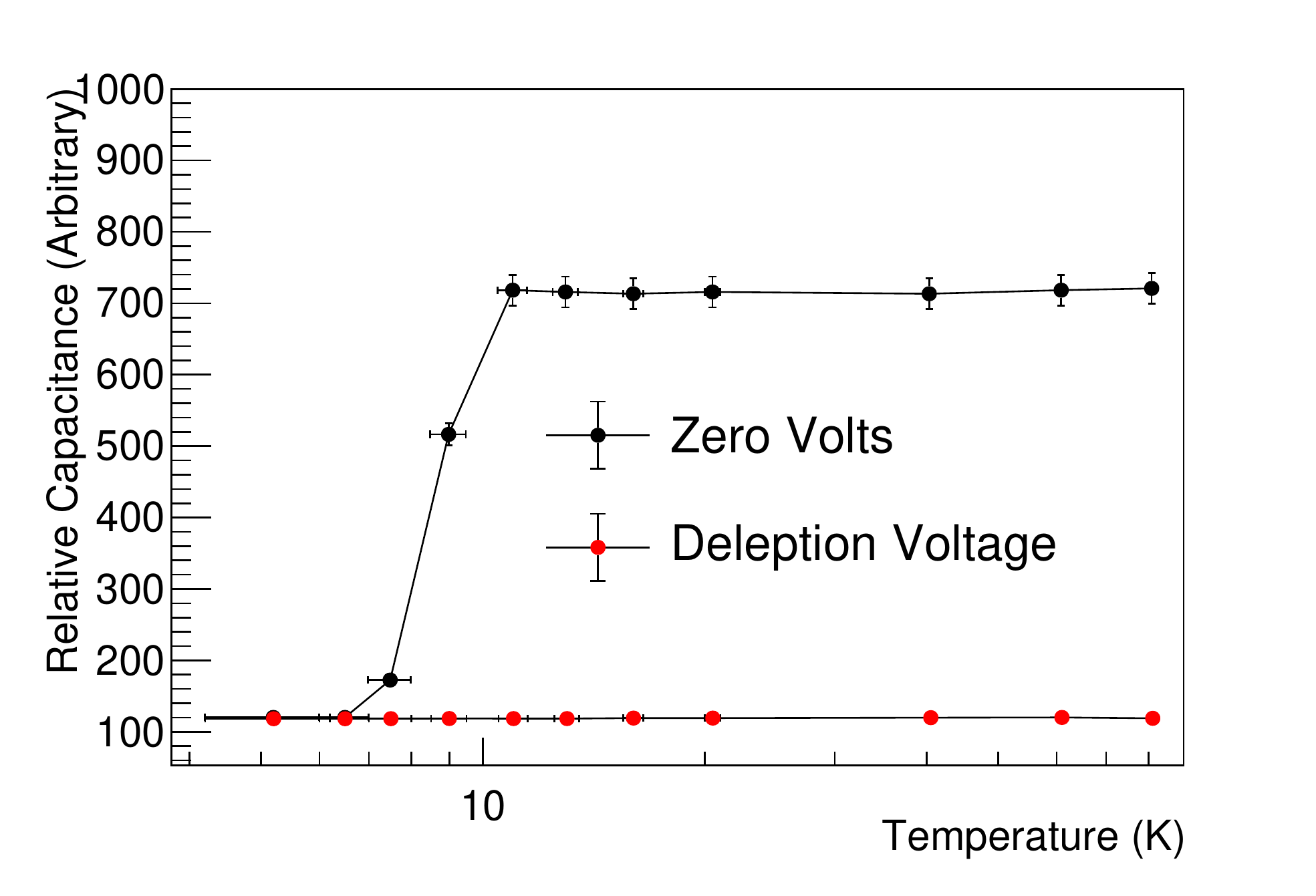}
  \caption{Shown is the relative capacitance versus temperature. The errors on the relative capacitance is within 3\% while the error of temperature is within 0.5 K. }
  \label{fig:f4}
\end{figure}

 Two detectors, an n-type (R09-02) and a p-type (RL) with respective net impurity densities of 7.02$\times$10$^{10}$/cm$^{3}$ and  6.2$\times$10$^{9}$/cm$^{3}$, are cooled down to 5.2 K. The dimensions of R09-02 are 11.7 mm $\times$ 11.5 mm $\times$ 5.5 mm while RL has dimensions of 18.8 mm $\times$ 17.9 mm $\times$ 10.7 mm.
 The relative capacitance of a given detector is measured at different temperatures as shown in Figure~\ref{fig:f4}. At zero bias, the relative capacitance stays constant until the temperature reaches about 10 K. Below 10 K, the relative capacitance decreases quickly to approach a constant level when the temperature reaches 6.5 K. Below this temperature, the relative capacitance is at the same level as the detector is fully depleted at 77.8 K. Similar behavior was observed in other early measurements~\cite{dven, dve1}. This demonstrates the detector is neutralized at a temperature below 6.5 K with zero bias voltage. In this temperature region, the formation of excited dipole states ensures the neutrality of the detector.  

The detector, R09-02 or RL, is biased according to a configuration with two operation modes: Mode 1: a negative bias is applied to the bottom surface; Mode 2: a positive bias is applied to the bottom surface. In both Mode 1 and Mode 2, the top surface is grounded and the charge signal is collected through the top contact. Using the localized charge carriers near the top surface created by $\alpha$ particles from $^{241}$Am, this configuration allows the electric field to drift one type of charge carriers across the detector at a given bias voltage. When the detector is operated in Mode 1, holes are drifted across the detector. While in Mode 2, electrons are drifted across the detector. Therefore, the charge trapping can be studied for holes and electrons separately for a given n-type or a p-type detector using $\alpha$ particles. For the widespread charge carriers created by the 661.7 keV $\gamma$ rays, both holes and electrons are drifted across the detector. Therefore, we expect to observe  a symmetrical charge collection  with respect to the polarity of the bias voltage. 

 An $^{241}$Am source is mounted above the detector (see Figure~\ref{fig:f3}.) inside the cryostat. The source is taken from a smoke detector so that $\alpha$ particles from the $^{241}$Am decays can arrive at the surface of the Ge detector. Since the sum of the Al layer and the a-Ge layer is very thin ($\sim$ 700 nm),  we observe the $\alpha$ peak in the spectra from the detector and perform different measurements over the course of two months. 
 \begin{figure}[htbp]
  \centering
  \includegraphics[clip,width=0.9\linewidth]{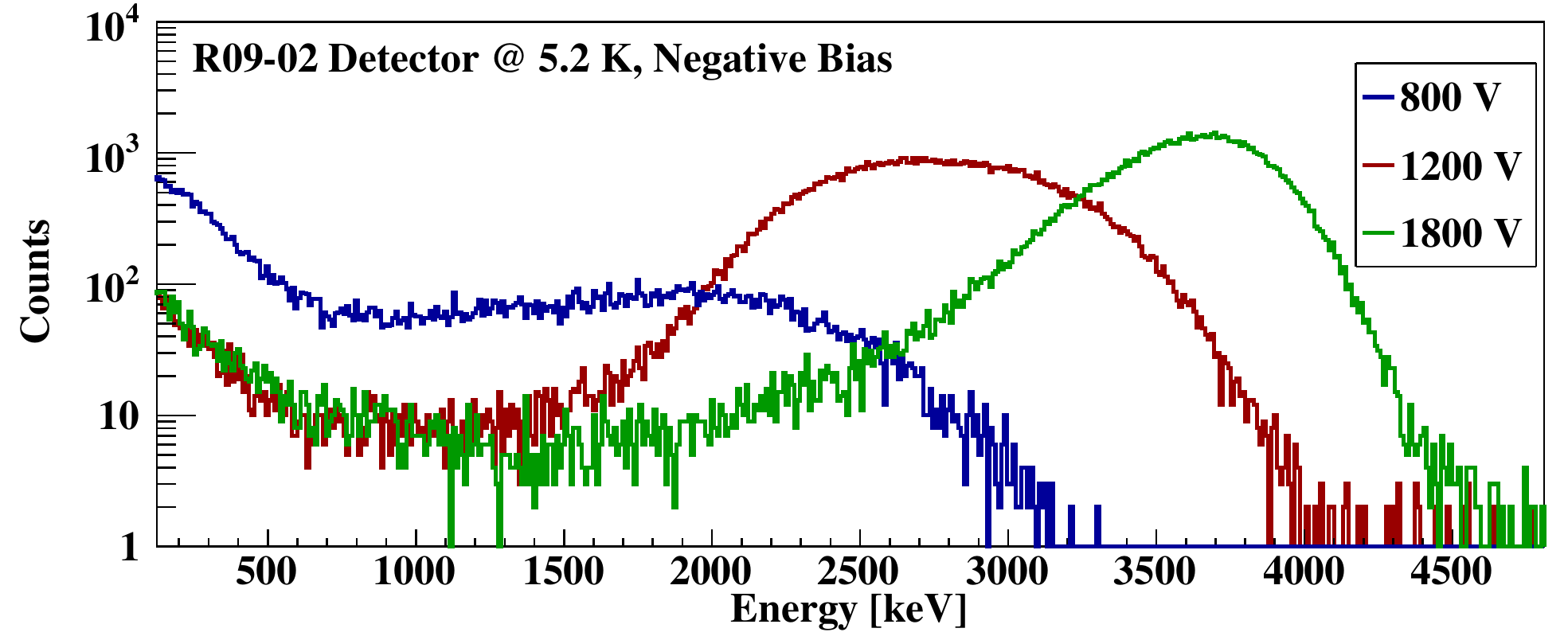}
  \includegraphics[clip,width=0.9\linewidth]{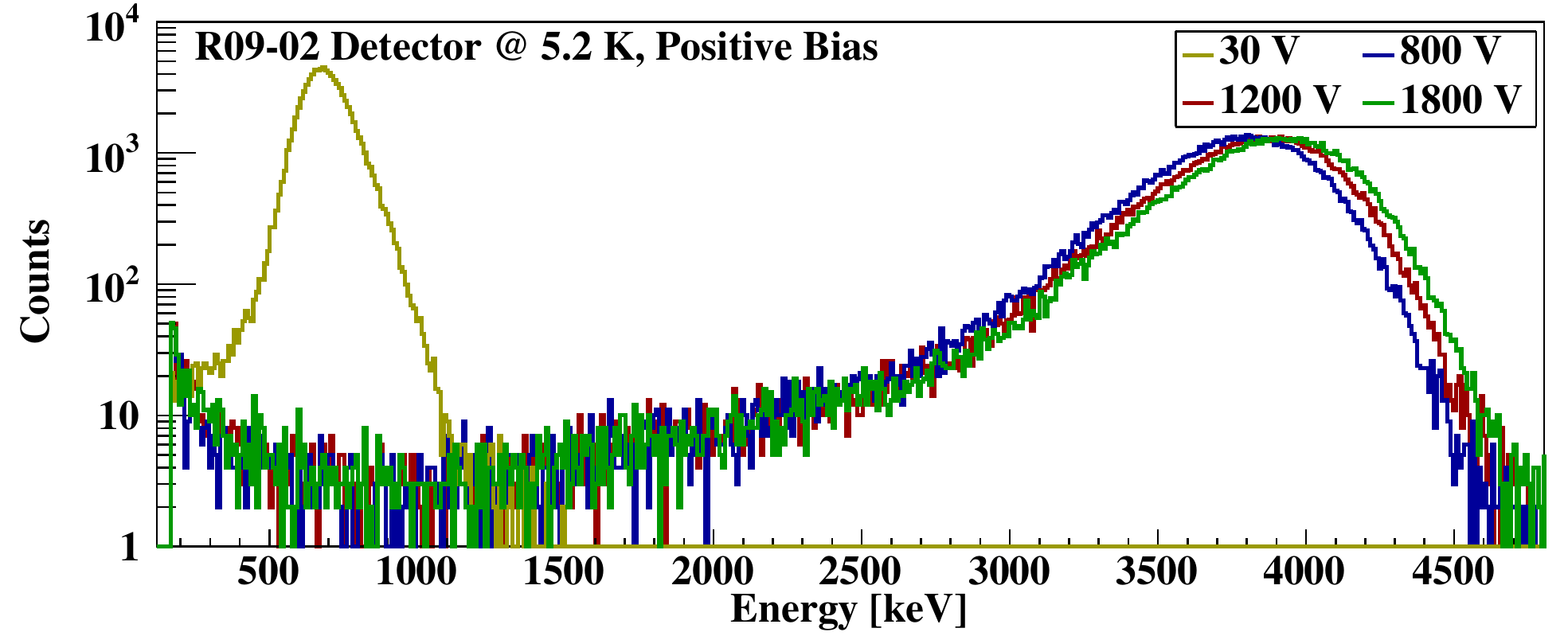}
  \includegraphics[clip,width=0.9\linewidth]{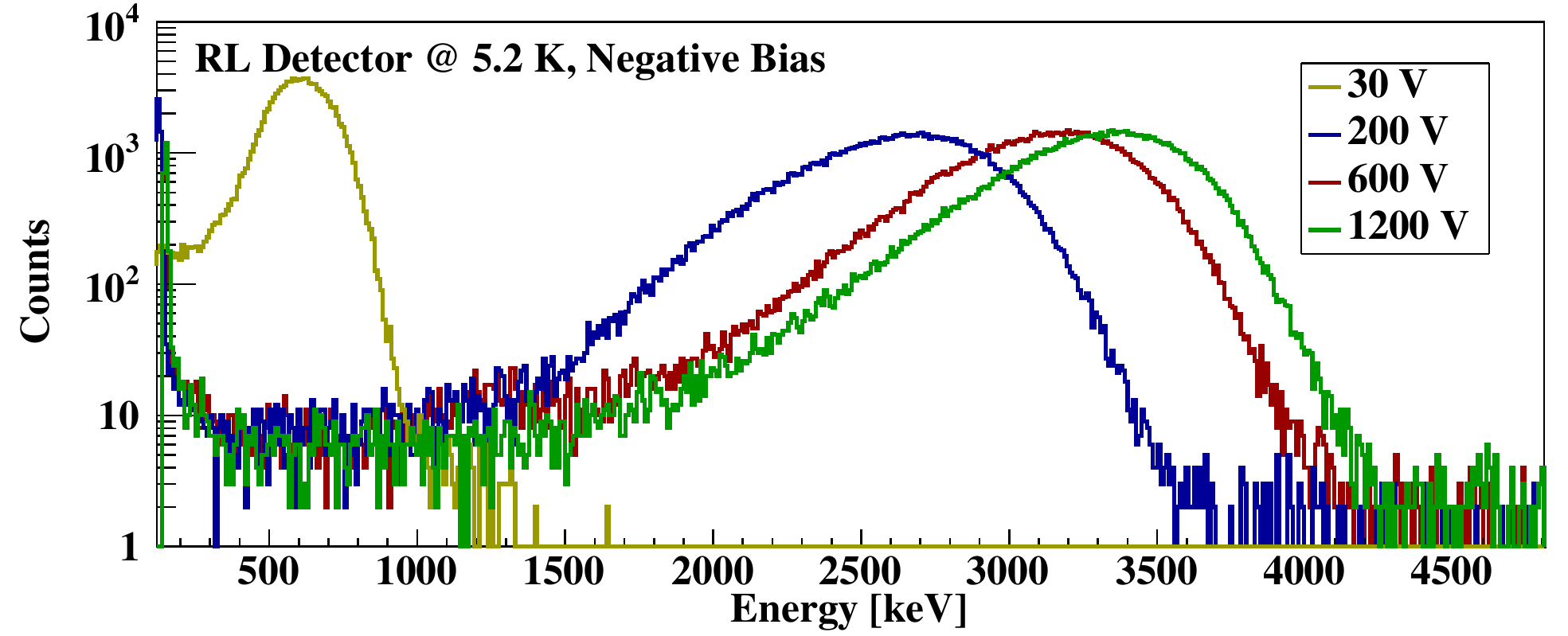}
  \includegraphics[clip,width=0.9\linewidth]{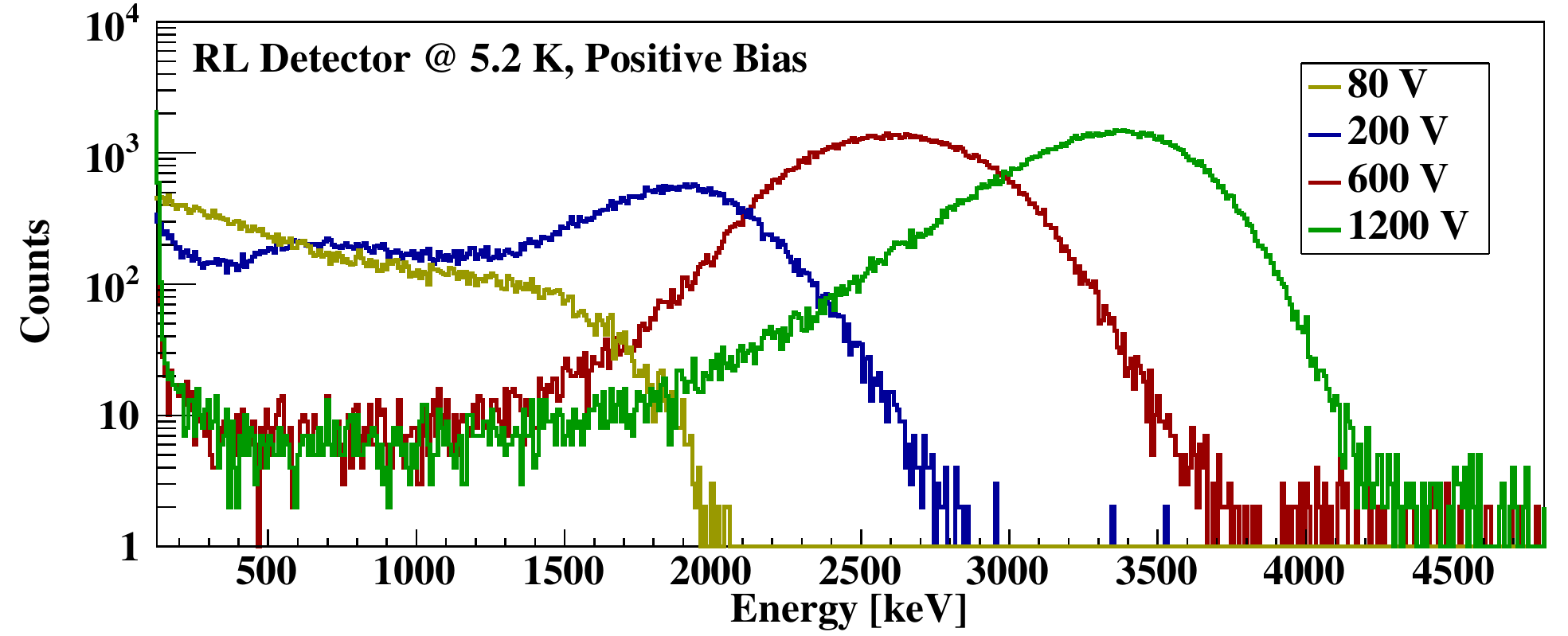}
  \caption{The first two plots show the measured energy deposition from the 5.3 MeV $\alpha$s in an n-type detector, R09-02. \#1: the detector was operated in Mode 1. \#2: the detector was operated in Mode 2. The second two plots show the measured energy deposition from the 5.3 MeV $\alpha$s in a p-type detector, RL. \#3: the detector was operated in Mode 1. \#4: the detector was operated in Mode 2.}
  \label{fig:f6}
\end{figure}
 
 We use the data measured at various bias voltages to characterize the charge trapping and to study how the charge trapping varies with bias voltage and the polarity of bias voltage. The first two plots (\#1 and \#2) of Figure~\ref{fig:f6} present the result of the 5.3 MeV $\alpha$ energy deposited in an n-type detector operated at 5.2 K. Since the $^{241}$Am source is positioned above the top of the detector, the localized holes produced on the surface are drifted through the full detector thickness in Mode 1 with a negative bias voltage.
 As can be seen from the first plot (\#1) of Figure~\ref{fig:f6}, the holes are significantly trapped when the detector is operated at lower bias voltages and the peak of 5.3 MeV is not even seen clearly until the bias voltage is increased to 1200 volts. The relative charge collection efficiency ($\varepsilon_h$) at 1200 volts is only $\sim$71\% when compared to 1800 volts. If one uses $\varepsilon_h = \frac{\lambda_h}{L}[1-exp(-L/\lambda_h)]$~\cite{mei1} to determine the trapping length for holes ($\lambda_h$) where $L$ = 5.5 mm is the detector thickness, $\lambda_h$ = 0.75 cm.  
 When the same detector is operated in Mode 2 with a positive bias applied to the bottom of the detector, electrons are drifted across the detector (\#2 of Figure~\ref{fig:f6}). The observation indicates a strong charge trapping at 30 volts, a factor of $\sim$7 reduction in charge collection, compared to 1800 volts. The charge collection efficiency ($\varepsilon_e$) at 1200 volts relative to 1800 volts is $\sim$99.9\%, which corresponds to a trapping length of 100 cm for electrons ($\lambda_e$) at 1200 volts. This observation suggests that electrons are less heavily trapped ($\lambda_e$ = 100 cm at 1200 volts) in an n-type detector when compared to holes ($\lambda_e$ = 0.75 cm at 1200 volts). The holes trapped by $D^{0^{*}}$ form $D^{+^{*}}$ states and the electrons trapped by $D^{0^{*}}$ form $D^{-^{*}}$ states in an n-type detector as demonstrated in Figure~\ref{fig:f1}. The observation shown in the first two plots (\#1 and \#2) of Figure~\ref{fig:f6} indicates the probability of forming $D^{+^{*}}$ states is larger than that of $D^{-^{*}}$ states. 
 
 Similarly, the energy deposition in the detector RL (a p-type detector) from the 5.3 MeV $\alpha$ particles is shown in the second two plots (\#3 and \#4) of Figure~\ref{fig:f6}. When the detector RL is operated in Mode 1 during which holes are drifted across the detector, a significant charge trapping, a factor of $\sim$7 reduction in charge collection at 30 volts compared to 1200 volts, is observed (\#3 of Figure~\ref{fig:f6}) . The charge trapping is also observed when the bias voltage is at 200 volts and 600 volts compared to 1200 volts. The charge collection efficiency ($\varepsilon_h$) at 600 volts relative to 1200 volts is $\sim$92\%, which corresponds to a trapping length of 6.5 cm for holes ($\lambda_h$) at 600 volts when $L$ = 10.7 mm. When the polarity of bias voltage is switched to Mode 2 during which electrons are drifted across the detector (\#4 of Figure~\ref{fig:f6}), more severe charge trapping is observed at 200 volts and 600 volts when compared to the same voltages in Mode 1. For example, the charge collection efficiency ($\varepsilon_e$) at 600 volts relative to 1200 volts is only $\sim$81.8\%, which corresponds to a trapping length of 2.6 cm for electrons ($\lambda_e$). This indicates that electrons trapped more severely ($\lambda_e$ = 2.6 cm at 600 volts) than holes ($\lambda_h$ = 6.5 cm at 600 volts) in a p-type detector. The electrons trapped by $A^{0^{*}}$ form $A^{-^{*}}$ states and the holes trapped by $A^{0^{*}}$ form $A^{+^{*}}$ states in a p-type detector as demonstrated in Figure~\ref{fig:f1}. The observation shown in the second two plots (\#3 and \#4 ) of Figure~\ref{fig:f6} suggests the probability of forming $A^{-^{*}}$ states is larger than that of $A^{+^{*}}$ states. 
 
 The observation of more holes (electrons) trapped in an n-type (a p-type) detector agrees with the prediction of the trapping mechanism: the Coulomb attraction induced by the excited dipoles to form cluster dipole states. This is because if the charge trapping were due to capture processes, one would expect electrons (holes) to be trapped more than holes (electrons) in an n-type (a p-type) detector since the same-side capture cross section (electrons capture by donors or holes capture by acceptors) is three orders of magnitude larger than the across band gap capture cross section (electron capture by acceptors and holes capture by donors)~\cite{phip1, phip, sund, sund1}.

\begin{figure}[htbp]
  \centering
  \includegraphics[clip,width=0.8\linewidth]{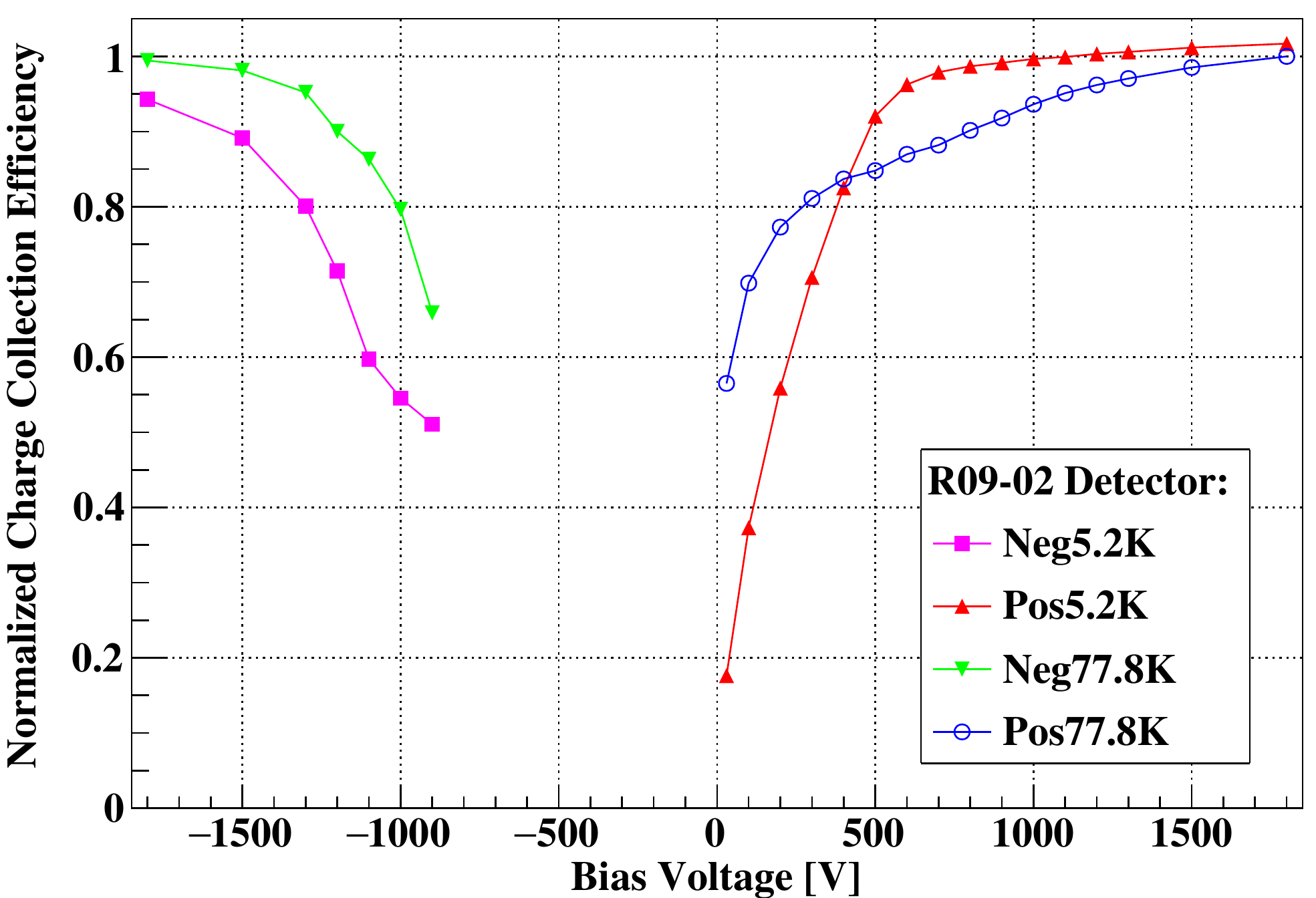}
  \includegraphics[clip,width=0.8\linewidth]{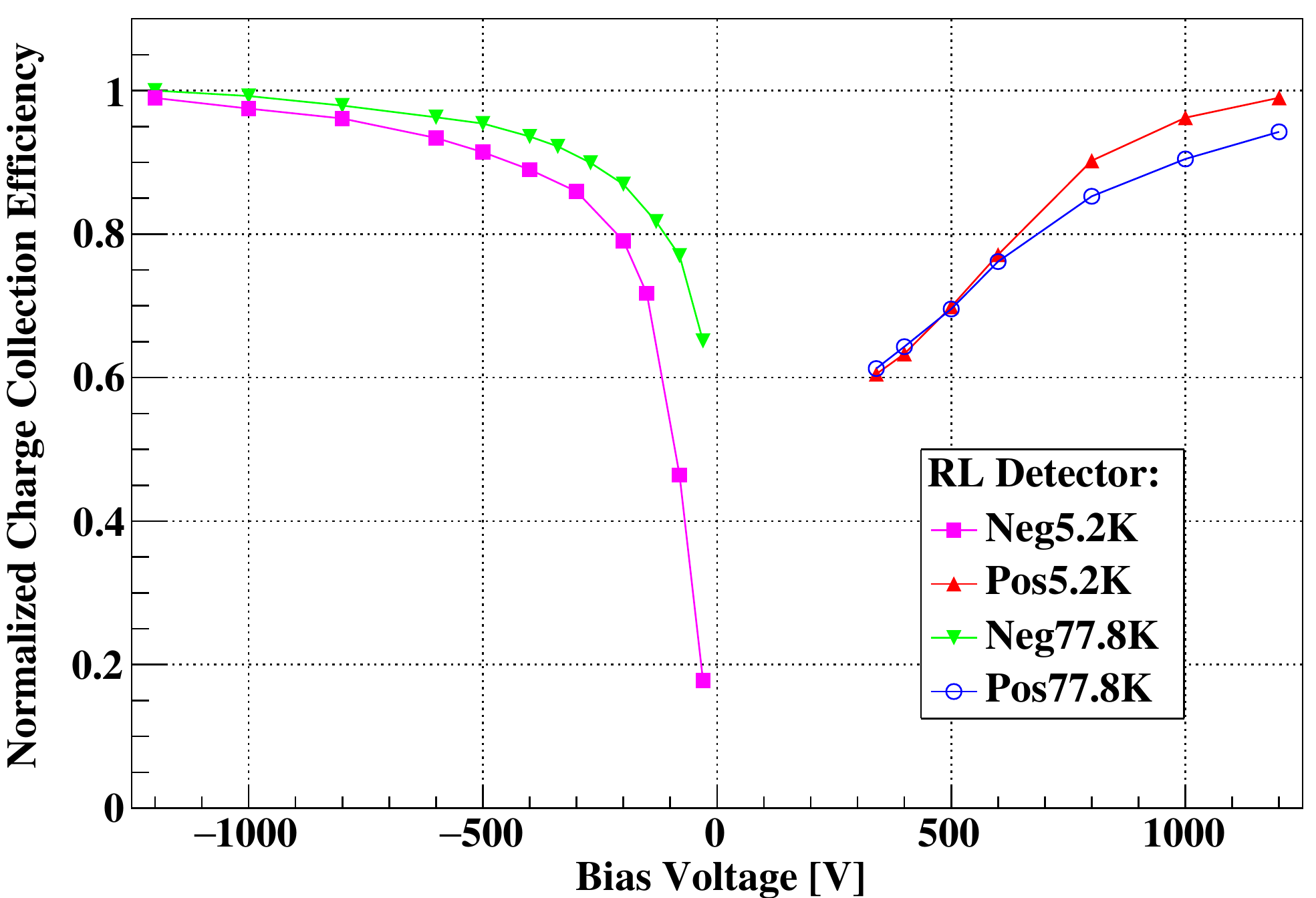}
  \includegraphics[clip,width=0.8\linewidth]{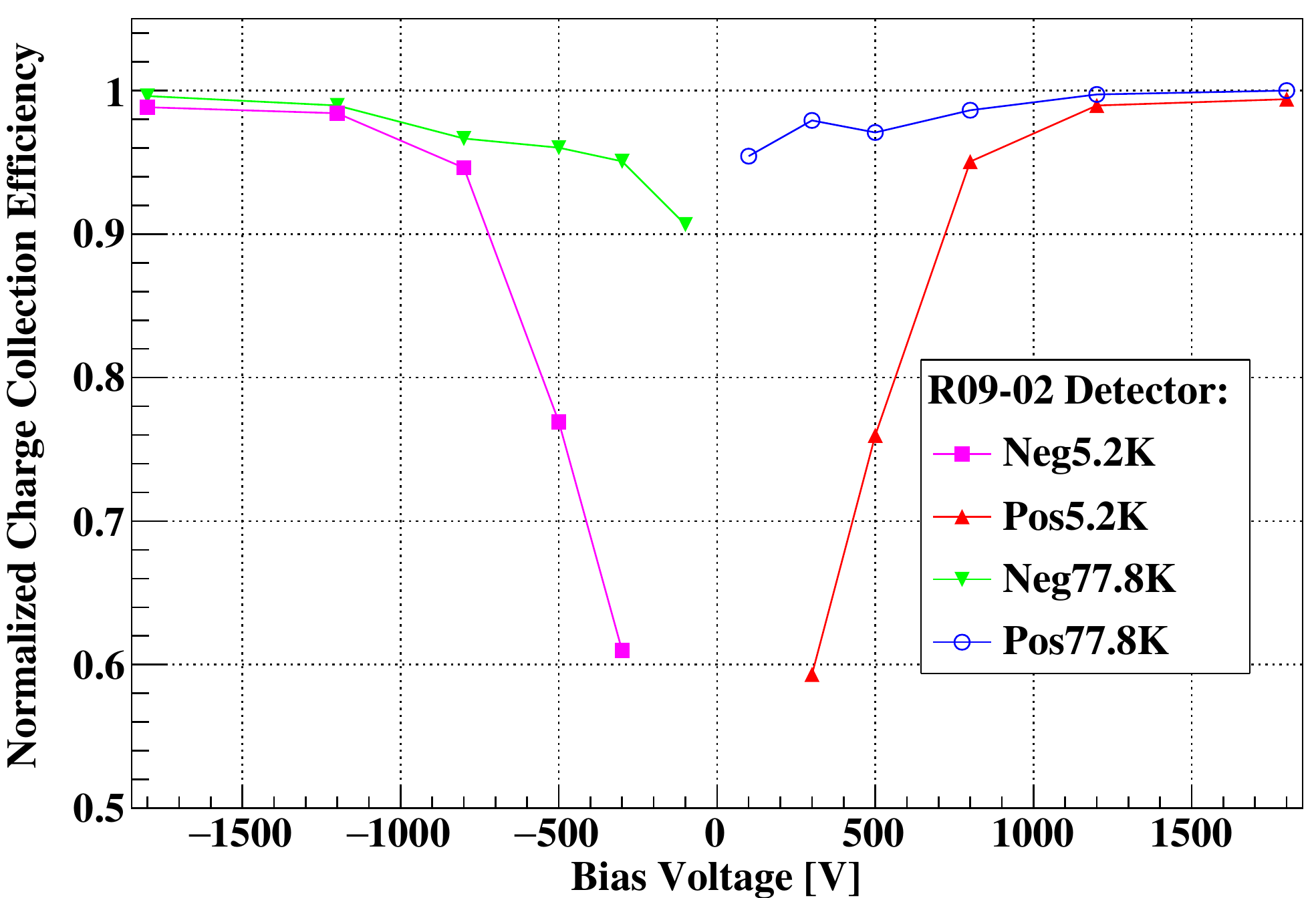}
  \includegraphics[clip,width=0.8\linewidth]{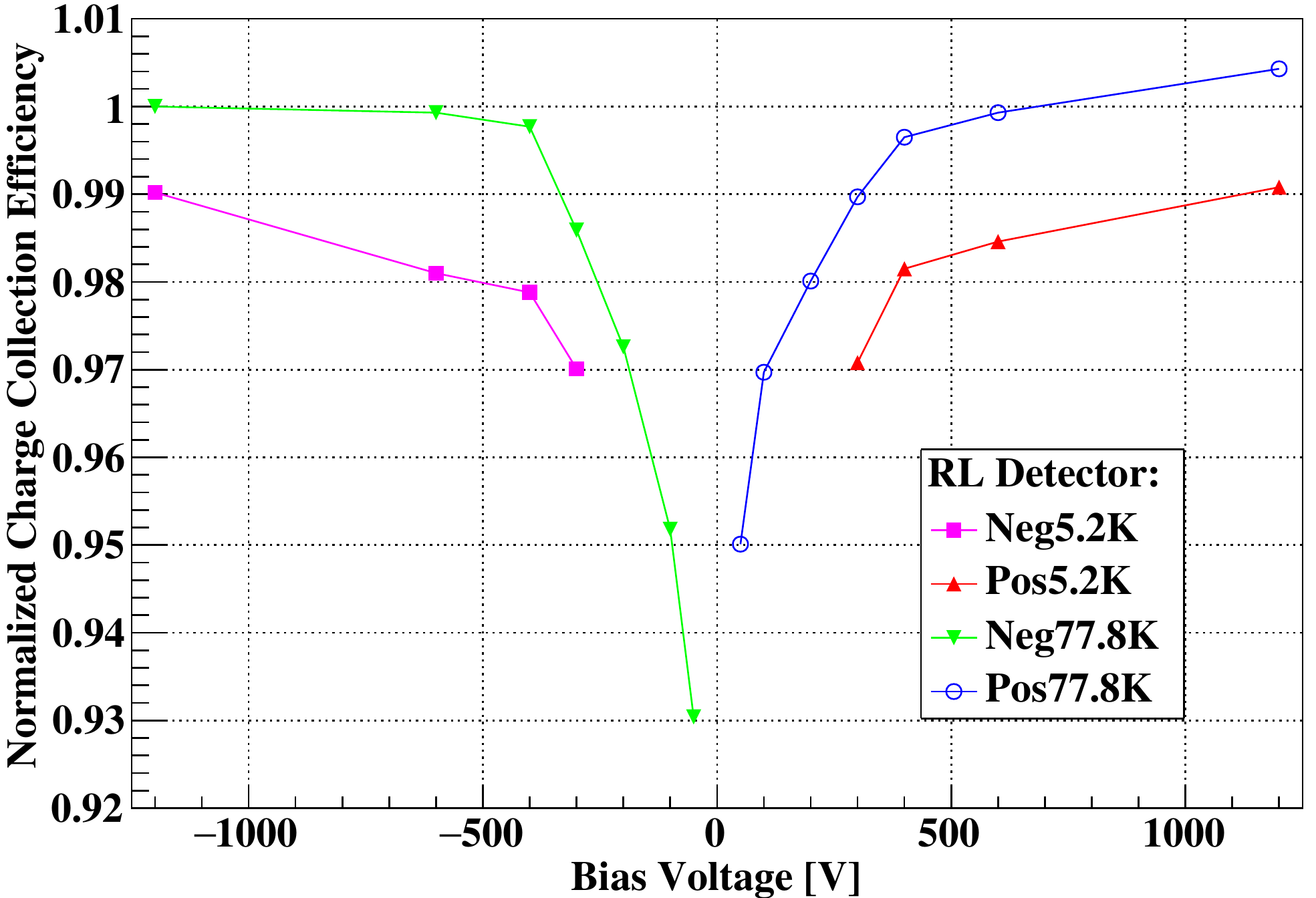}
  \caption{The first two plots (\#1 and \#2) show the normalized charge collection for the charge created by the 5.3 MeV $\alpha$ particles versus the applied bias voltage at different temperatures with Mode 1 and Mode 2. \#1: R09-02, an n-type detector. \#2: RL, a p-type detector. The second two plots (\#3 and \#4) show the normalized charge collection efficiency for the charge created by the 661.7 keV $\gamma$ rays versus the applied bias voltage at different temperatures with Mode 1 and Mode 2. \#3: R09-02, an n-type detector. \#4: RL, a p-type detector.}
  \label{fig:f8}
\end{figure}

 As a cross check of the charge trapping mechanism, both detectors (R09-02 and RL) are operated at 77.8 K at which the charge trapping is dominated by the space charged-states (donor ion states and acceptor ion states) left from depleting the detector. The first two plots (\#1 and \#2) of Figure~\ref{fig:f8} show the normalized charge collection efficiency for the charge created by the 5.3 MeV $\alpha$ particles versus the applied bias voltage at different temperatures with Mode 1 and Mode 2. The normalization is performed for both detectors operated at 77.8 K with 1800 volts for R09-02 detector and -1200 volts for RL detector. 
 At 77.8 K, the detector R09-02 is fully depleted at 1200 volts. When it is operated at 1800 volts, charge trapping is minimized and the peak position of $\alpha$ is determined to be $\sim$3.8 MeV, which implies an energy loss of 1.5 MeV due to continuous energy loss of $\alpha$ particles on the way from the source to the detector. The energy scale of the detector is calibrated by well known $\gamma$ rays. Similarly, the detector RL is fully depleted at 400 volts. When the detector is operated at 1200 volts, the peak position of $\alpha$ is determined to be $\sim$3.3 MeV. The difference in energy loss of $\alpha$ particles on the way from the source and the detector between R09-02 and RL is mainly due to the thickness of the a-Ge layer as the electrical contacts for both detectors. The difference in the peak position of $\alpha$ at different bias voltages relative to the peak position at 77.8 K operated at 1800 volts for R09-02 and 1200 volts for RL demonstrates the amount of charge trapping. At 77.8 K, charge trapping is dominated by the Coulomb attraction between the space-charged states and the corresponding charge carriers. At 5.2 K, since the detector is neutralized in charge states, charge trapping is mainly due to the Coulomb attraction between dipole states and the corresponding charge carriers. 
 
 To further verify the charge trapping mechanism, both detectors are exposed to the 661.7 keV $\gamma$ rays from a $^{137}Cs$ source. Both positive bias and negative bias are applied to the bottom surface of both detectors. Since widespread charge carriers inside the detector are expected from the 661.7 keV $\gamma$ rays, the charge trapping is anticipated to have no dependence on the polarity of bias voltage  in both detectors. The second two plots (\#3 and \#4) of Figure~\ref{fig:f8} show the normalized charge collection efficiency versus the applied bias voltage for both detectors at different temperatures. As can be seen, the amount of charge trapping between positive bias and negative bias in both detectors is almost the same. The charge trapping has no dependence on positive or negative bias voltages in both detectors. The correlation between charge trapping and the impurity level is noticeable from the measurements of the 661.7 keV $\gamma$ rays in R09-02 and RL detectors operated at 5.2 K. For example, when both detectors are applied a bias of 300 volts, more charge trapping (a factor of $\sim$1.6 reduction in charge collection)  is observed in R02-02 detector (\#3 of Figure~\ref{fig:f8}) compared to the charge trapping in RL (\#4 of Figure~\ref{fig:f8}), which has a smaller net impurity concentration.  It is worth mentioning that significant charge trapping at 4.2 K was also reported by D. V\'{e}nos et al.~\cite{dven, dve1} in p-type detectors operated at lower bias voltage.

We conclude that the observed charge trapping at 5.2 K is due to the Coulomb attraction between excited dipole states and corresponding charge carriers. At 5.2 K, it is tempting to speculate that cluster dipole states are being formed and this is consistent with the theory of Abakumov~\cite{abak1, abak2}. The formation of cluster dipole states is responsible for the observed significant charge trapping at a lower voltage. Using both n-type and p-type detectors, we find that holes are trapped more in an n-type detector and electrons are trapped more in a p-type detector at 5.2 K. This phenomenon is attributed to a larger probability of forming $D^{+^{*}}$ (or $A^{-^{*}}$) states than that of $D^{-^{*}}$ (or $A^{+^{*}}$) states in an n-type (or a p-type) detector.  These cluster dipole states are thermally stable with a binding energy, which is greater than $k_{B}$T at a level of 0.45 meV at 5.2 K. The amount of charge trapping as a function of the external field can be related to the charge trapping length, charge trapping cross sections, and the binding energy of the excited dipole states, which is beyond the scope of this work and will be reported in the follow-up papers.
 
 The data that supports the findings of this study are available within the article. 
 The authors would like to thank Christina Keller for a careful reading of this manuscript. This work was supported in part by NSF OISE 1743790, DE-SC0004768, and the State of South Dakota via a research center. 
 

\end{document}